\documentclass[apl,reprint,twocolumn,floatfix,superscriptaddress]{revtex4-1}
\usepackage[english]{babel}
\usepackage{graphicx}
\usepackage{verbatim}

\usepackage{epsfig}
\usepackage{epstopdf}
\usepackage{amsfonts}
\usepackage{amsthm}
\usepackage{amsmath}
\usepackage{amssymb}
\usepackage{hyperref} 
\usepackage[usenames,dvipsnames,svgnames,table]{xcolor}
\hypersetup{colorlinks=true,linkcolor=NavyBlue,citecolor=BrickRed,urlcolor=BrickRed}
\usepackage{makeidx}
\usepackage{pifont}
\usepackage{pdfpages}

\usepackage{color}
\usepackage{grffile}


\def\expect#1{\mathinner{\langle{#1}\rangle}}

{\catcode`\|=\active
  \gdef\expect#1{\left<\mathcode`\|"8000\let|\bravert {#1}\right>}}
\def\bravert{\egroup\,\vrule\,\bgroup}

\def\beq{\begin{equation}}
\def\eeq{\end{equation}}
\def\be{\begin{equation}}
\def\ee{\end{equation}}

\def\cG0{{\cal G}_0}

\def\q{{\bf q}}

\def\a{\alpha}

\def\D{\Delta}

\def\G{\Gamma}

\def\uc2{$U_{c2}$}
\def\uc1{$U_{c1}$}

\def\bea{\begin{eqnarray}}
\def\eea{\end{eqnarray}}
\def \bal{\begin{align}}
\def \eal{\end{align}} 
\def\#{\!\!}
\def\@{\!\!\!\!}

\def\+{\dagger}

\begin{document}

\title{A thermomagnetic mechanism for self-cooling cables}
\author{Luca de'~Medici}
\affiliation{European Synchrotron Radiation Facility, 71 Av. des Martyrs, Grenoble, France}
\affiliation{Laboratoire de Physique et Etude des Mat\'eriaux, UMR8213 CNRS/ESPCI/UPMC, Paris, France}

\date{\today}

\begin{abstract}
A solid state mechanism for cooling high-current cables is proposed based on the Ettingshausen effect, i.e. the transverse thermoelectric cooling generated in magnetic fields. 
The intense current running in the cable generates a strong magnetic field around it, that can be exploited by a small current running in a coating layer made out of a strong "thermomagnetic" material to induce a temperature difference between the cable core and the environment. 
Both analytical calculations and realistic numerical simulations for Bismuth coatings in typical magnetic fields are presented. The latter yield temperature drops $\simeq$60K and $>$100K for a single- and double-layer coating respectively.
These encouraging results should stimulate the search for better thermomagnetic materials, in view of applications such as self-cooled superconducting cables working at room temperature.
\end{abstract}

\maketitle

Transmission of intense electric currents through conducting cables is of obvious importance for energy supply. However the present distribution through high-voltage power-lines suffers considerable losses due to Joule effect caused by the resistance of the metallic cables.
A promising alternative is represented by superconducting cables, where the resistance is zero. However present-day materials reach the superconducting state below very low temperatures ($T_c \simeq 18K$ for a widespread "conventional" superconductor, Nb$_3$Sn, while $T_c \simeq 90K$ for one of the new "high-temperature" superconductors, YBCO), making their technological use viable only in specific conditions in which cooling to cryogenic temperatures is possible.
Both these types of cables are also used in the coils of high-field magnets used in medical applications like magnetic resonance imaging, in levitating train technology, energy storage and for scientific purposes.

In all these situations a solid-state method for cooling the cables can potentially improve the performances by lowering the resistance in resistive cables and helping or even replacing the existing cryogenic methods to lower the environmental temperature of superconducting cables.
I propose such a potentially useful method here.
\begin{figure}[h!]
\begin{center}
\includegraphics[width=6cm]{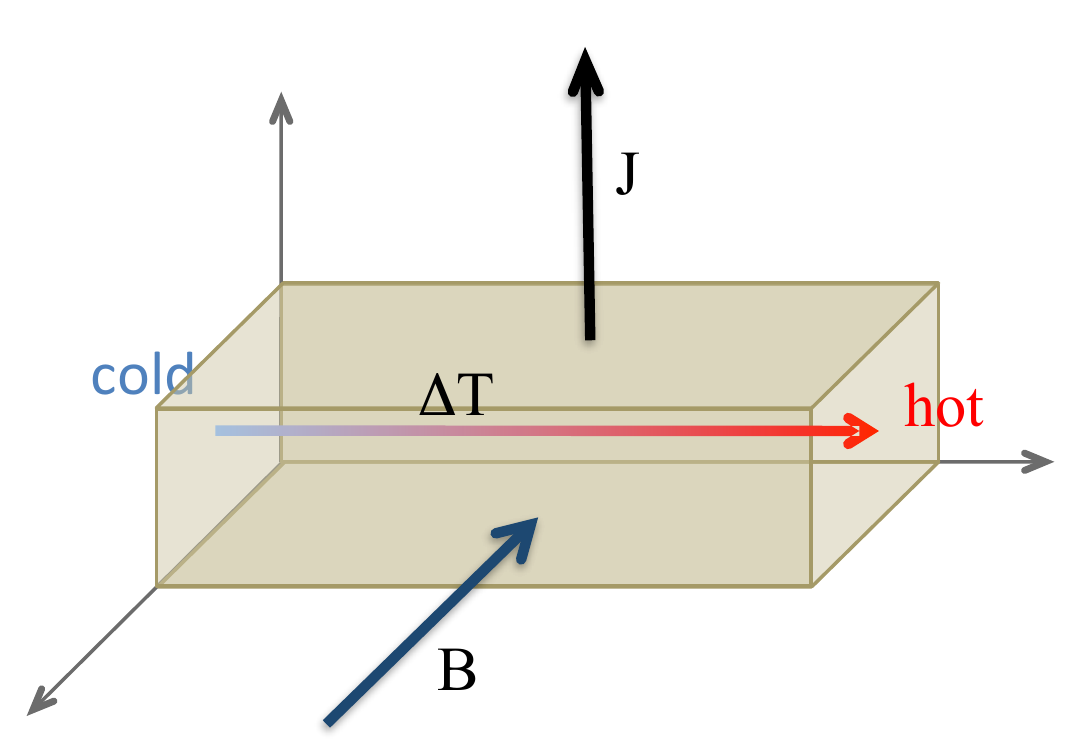}
\includegraphics[width=5.cm]{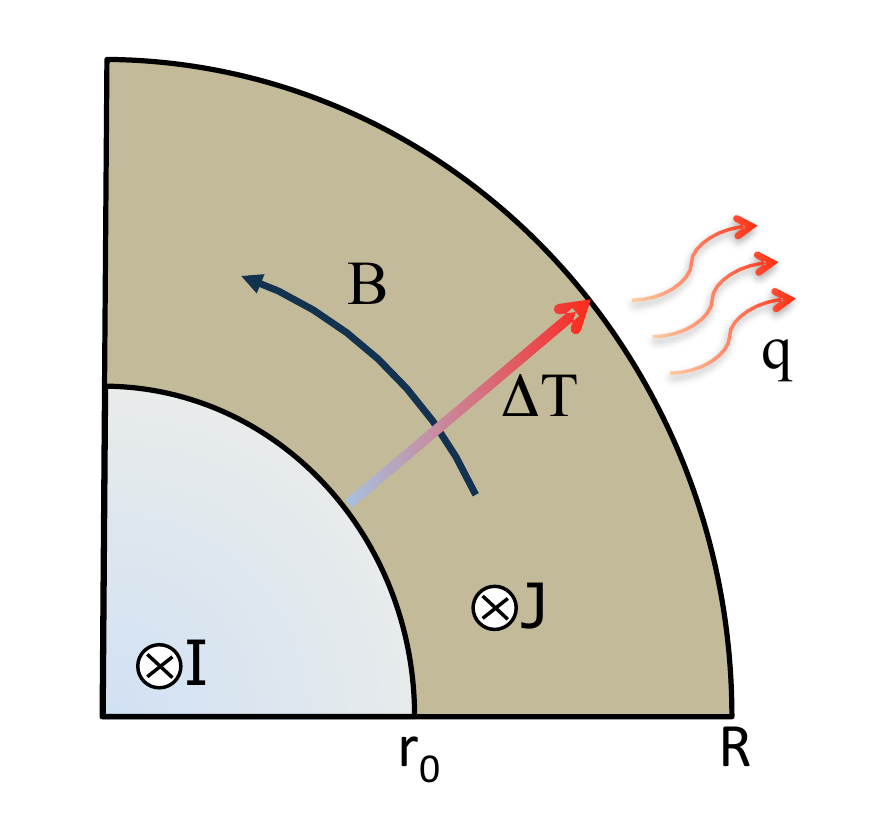}
\caption{{\bf Upper panel}: illustration of the Ettingshausen effect in a slab of rectangular section. A current flowing in the $\hat z$ direction in presence of a magnetic field oriented in the $\hat y$ direction generates a temperature gradient in the $\hat x$ direction. {\bf Lower panel}: sector of a section of the coaxial cable in the proposed setup. The (super)conductor carrying the main current is depicted in grey ($r<r_0$), the thermomagnetic material in beige ($r_0<r<R$). Electrical insulation is implied between the layers. The magnetic field $B$ is generated by the main current I, whereas the temperature gradient is the outcome of the Ettingshausen effect due to the current J.}\label{fig:Ettingshausen-cable}
\end{center}
\end{figure}

In the proposed setup the cable transporting the main current is coated with - but electrically insulated from - a layer of another material with strong thermomagnetic properties. 
A "thermomagnetic material", like elemental Bismuth, is one that manifests strong Nernst-Ettingshausen effect, which is the transverse counterpart of thermoelectric (Seebeck-Peltier) effects induced by a strong magnetic field.
In particular the Ettingshausen effect (see the upper panel of Fig. \ref{fig:Ettingshausen-cable}), is the generation of a temperature gradient in the direction orthogonal to both the electric current flowing in a material and the applied perpendicular magnetic field.
The technological use of thermoelectricity (for coolers, waste heat recoverers, nuclear generators in space missions, etc.), utterly boosted by the recent improvement in materials performances due to a strong research effort, is already a reality, and growing\cite{Rowe-Handbook_Thermoelectrics}.
Thermomagnetic effects instead have been much more neglected for technological developments due to the impracticality of incorporating strong permanent magnets into devices.

What I point out in this article (see the lower panel of Fig.\ref{fig:Ettingshausen-cable}) is that in a high-power cable the main current $I$ transported in the core generates a magnetic field around it that can be used for Ettingshausen cooling of the cable itself. Indeed this field is in the right geometry for an auxiliary current $J$ running into the outer thermomagnetic layer to induce a temperature gradient between the cable core and the outer environment.
The two current should not be confused: the density of the main current will be typically several orders of magnitude larger than that of the auxiliary current.

\medskip

{\bf \footnotesize \noindent MAGNITUDE OF THE EFFECT}

\noindent One may rightfully wonder if such an effect can be large enough for practical use. Indeed one can consider two possible applications of extreme technological interest, among others, like cooling a cable with a high-$T_c$ superconductor core for it to function in an environment at ambient temperature, or for a cable of conventional superconductor to function at liquid Nitrogen temperature. For these situations very large temperature drops induced by the cooling mechanism are needed, with T$_l <$ T$_h/2$ (where T$_l$ and T$_h$ are the temperatures of the cold side and of the hot side of the cooling coating, respectively, such that $\D T=T_h-T_l$ the temperature drop due to the Ettingshausen cooler). 
Bismuth and its Sb-alloys are known for having the strongest Nernst-Ettingshausen effect with a thermomagnetic power $NB$ largely surpassing the mV/K  in a range of temperatures\cite{Behnia-Bismuth, Cuff-BiSb_Ettingshausen_properties_cooling} (where $N(T,B)$ is the Nernst coefficient defined as the voltage generated in a Nernst setup per Kelvin per Tesla, and B the magnetic field). The performance of a cooling device is however not only set by the thermomagnetic power, but also by the material electrical resistivity $\rho(T,B)$ and thermal conductivity $K(T,B)$. Indeed a high thermal conductivity eases the heat back-flow and makes the task of maintaining the temperature gradient harder.
Moreover the heat produced by the Joule effect contrasts the cooling mechanism in general, thus a high resistivity is detrimental to a good performance. It is to be noted that the Joule heating grows quadratically with the electric current, while the Ettingshausen effect only linearly. This implies that there is an optimal applied voltage for which an Ettingshausen device reaches the maximal temperature difference $\D T$, beyond which $\D T$ decreases and vanishes.

In practice the combination of the above quantities that better describes the thermomagnetic performance is the \emph{figure of merit} $Z=\frac{(NB)^2}{\rho K}$, which is indeed also a function of temperature and magnetic field. It has the dimension of an inverse temperature so that it is customary to characterize a material by its adimensional figure of merit ZT, which can assume values between 0 and 1. Measured ZT for Bi and BiSb alloys reaches, in high magnetic fields, values between 0.2 and 0.4 in the temperature range 100-300K\cite{harman-ZNE_Bismuth} (inset of Fig. \ref{fig:Cable_Bi}). 
All the above quantities are intended defined in a transversely isothermal setup (see supplementary information). Another useful quantity is the adiabatic figure of merit $Z_a$, defined analogously to Z, but using the adiabatic resistivity. The two figures of merit are related by $Z_a=\frac{Z}{1-ZT}$.

It can be shown\cite{OBrien_Wallace,HarmanHonig-I,kooi-Ettingshausen_cooler} that in a parallalelepiped setup like the one depicted in the upper panel of Fig.\ref{fig:Ettingshausen-cable} the expected temperature drop $\D T$ is roughly (exactly, if $N$, $\rho$, $K$ and thus $Z$ are independent of temperature):
\be\label{eq:slab_DTmax}
\D T_{MAX}=\frac{1}{2}Z T^2_h=\frac{1}{2}Z_a(\bar T) T^2_l
\ee
where $\bar T=(T_h+T_l)/2$ is the middle temperature between the hot and cold side.
$\D$T is then for example of the order of 30K for a material with ZT=0.2 in a device with a heat sink at the ambient temperature of 300K. 
Indeed experimental measures on blocks of good thermomagnetic materials have been performed in similar conditions: Kooi et al.\cite{kooi-Ettingshausen_cooler, Cuff-BiSb_Ettingshausen_properties_cooling} report, in accordance with eq.(\ref{eq:slab_DTmax}), a measured $\D $T$\simeq$30K for a rectangular block of single-crystal Bi(97)Sb(3) in a magnetic field of 1T with the hot side temperature T$_h$=156K (around which ZT$\sim$0.4) and an average current density of 100 A/cm$^2$.

The theoretical framework leading to formulas (\ref{eq:slab_DTmax}) is strictly analogous to that of Peltier cooling (with the adiabatic thermomagnetic figure of merit $Z_a$ replacing the thermoelectric figure of merit\cite{kooi-Ettingshausen_cooler,goldsmid_book2009}). In Peltier coolers both the efficiency and $\D T_{MAX}$ can be substantially improved if a "cascade" of cooling devices is used, in which every stage is the heat sink of the previous one and has a larger cooling power\cite{goldsmid_book2009}. In Ettingshausen coolers one can simply shape the cooling block in a way that the heat sink side is larger than the heat source side in order to continuously increase the cooling power within a single stage. Thus a shape (a trapezoidal block with exponential sides) corresponding to "infinite staging" has been obtained\cite{OBrien_Wallace,delves1962prospects,harman-InfiniteStaged_theory,scholz1994infinite} that was capable to reach $\D$T=101K for a single crystal Bi block with a heat sink face 128 times larger than the heat source face in $\simeq$11T magnetic field\cite{harman-shaping101K}.

In the cooling setup proposed in this paper some shaping is present simply due to the circular shape, that implies a heat sink face larger than the heat source face, and one can expect better performances than in an analogous cooler with the shape of a parallelepiped\cite{kooi_staging}. Moreover staging, i.e. multiple thermomagnetic layers with different currents $J_{1}, J_{2},\ldots$ and/or materials can be still used to further improve them.

It is to be noted that $\sim 10T$ is the order of magnitude of the magnetic field just outside high-current cables used in both resistive and magnetic coils. Cables generating higher fields can also be considered since the upper critical fields of both conventional and high-Tc superconductors are even higher (for Nb$_3$Sn $H_{c2}\simeq 30T$, for YBCO $H_c2\simeq 120T$, for the recently discovered Fe-superconductors\cite{Springer_book_FeSC} $H_c2\gtrsim 50T$).

\medskip

{\bf \footnotesize \noindent THE FUNDAMENTAL EQUATIONS FOR A CIRCULAR ETTINGSHAUSEN COOLER}

\noindent In order to evaluate quantitatively the performances of the proposed circular Ettingshausen setup one has to consider the transport equations for thermoelectric effects in a magnetic field that can be derived in the linear response approximation using out-of-equilibrium thermodynamics\cite{callen2006thermodynamics}. 
We consider a cylindrical cable (the axis of the cable is taken to be in the $\hat z$ direction) with a core of radius $r_0$ and a coating layer of external radius R, as in the lower panel of Fig. \ref{fig:Ettingshausen-cable}. The cable is much longer than wide, so that besides some effects at the ends of it, all quantities are constant by respect to $z$ and $\theta$ and the heat flows radially.
Following Kooi et al.\cite{kooi-Ettingshausen_cooler} the relevant equations in the present case, written in cylindrical coordinates, are:
\bea
\label{eq:Jz} j_z&=\#&\frac{E_z}{\rho}+\frac{NB}{\rho}\frac{dT}{dr},\\
\label{eq:qr} q_r&=\#&\frac{NBT}{\rho}E_z+K(ZT-1)\frac{dT}{dr},
\eea
where $j_z$ and ${E_z}$ are respectively the electric current density and the electric field, along the direction of the cable, within the thermomagnetic layer (${\mathbf E}=-{\mathbf \nabla} \mu$, where $\mu$ is the electrochemical potential),  whereas $q_r$ is the heat current density in the radial direction and $T(r)$ is the absolute temperature.  

All quantities are function of $r$, even if the continuity of the electrochemical potential and the symmetry-imposed constant value of $E_r$ as a function of z imply that $E_z$ is also a constant in r (because ${\partial E_z}/{\partial r}=-\frac{\partial^2 \mu}{\partial r \partial z}={\partial E_r}/{\partial z}=0$)\cite{kooi-Ettingshausen_cooler}. Eq.~(\ref{eq:Jz}) shows then that the current density $j_z$ can vary with r.
$B(r)= B_0r_0/r$ is the intensity of the magnetic field (oriented along $\hat \theta$), which we parametrize by its value at the interface between the conducting core and the thermomagnetic coating 
$
B(r=r_0)=B_0=\frac{\mu_0}{2\pi}\frac{I}{r_0}=\frac{\mu_0}{2}r_0i,
$  
where $\mu_0$ is the vacuum permittivity and $I$ and $i$ are the total current and current density in the cable core, respectively
\footnote{In principle one should also consider the magnetic permittivity of the thermomagnetic material $\mu_r$ in the calculation of B. Bismuth is a diamagnetic material in which diamagnetism is particularly relevant (and the first discovered), however it is still negligible in practice: $\chi_m=\mu_r-1=-1.66\cdot 10^{-4}$, i.e. $\mu_r=0.999834$}.

In order to find the temperature and current distribution, energy conservation has to be enforced at each point by a continuity equation\cite{kooi-Ettingshausen_cooler} $\nabla \cdot ({\mathbf q} + \mu {\mathbf j})=0$, that here specializes to:
\be\label{eq:continuity}
\frac{1}{r}\frac{d\;\: rq_r}{dr}=E_zj_z
\ee

The set of unidimensional equations (\ref{eq:Jz}), (\ref{eq:qr}) and (\ref{eq:continuity}) can be solved numerically, especially when a material specific dependence in T and B of the transport coefficients has to be taken into account\cite{scholz1994infinite}. However we can gain some physical insight analytically first, in particular cases.

\medskip

{\bf \footnotesize \noindent PHYSICAL INSIGHT THROUGH AN ANALYTICAL MODEL}

\noindent Indeed it is natural to consider the case of constant $\rho$, $N$ and $K$, to be compared to the analogous treatment for a rectangular cooler performed in Ref. \onlinecite{kooi-Ettingshausen_cooler}. 
In this case equations (\ref{eq:Jz}), (\ref{eq:qr}) and (\ref{eq:continuity}) can be solved analytically (with a minor additional approximation in the present case, see the supplmentary material) for the temperature and heat flow distributions $T(r)$ and $q_r(r)$.

These solutions are parametrized by the boundary conditions such as $T(R)=T_h$, and $q_r(r_0)=q_{IN}$, so that, for a given material, the electric field is the only remaining knob, and the inner temperature $T_l=T(r_0)$ is then a function of $T_h$, $q_{IN}$ and $E_z$.
For a cooling device coating a superconducting core (which does not produce any heat in the conduction) $q_{IN}=0$ in the steady state, whereas for the device coating a resistive core $q_{IN}$ is imposed and finite. Also for all stages beyond the first in a multi-stage setup $q_{IN}$ is finite.

We are interested here in knowing what is the maximum temperature drop that one can obtain for a given $q_{IN}$. As said there is an optimal value of the electric field beyond which the temperature drop recedes. This is obtained by minimizing $T_l$ at fixed $q_{IN}$ and is:
\be\label{eq:Emax}
E_z^{MAX}=\frac{NB_0T_l}{r_0\G_\a},
\ee
where $\G_\a=\frac{\a^2-1}{2\log{\a}}-1$, and $\a=R/r_0$ measures the thickness of the thermomagnetic layer ($\a>1$).
This formula is very similar to that for a rectangular cooler\cite{kooi-Ettingshausen_cooler} (where $E_z^{MAX}={NBT_l}/{b}$, and $b$ is the thickness of the cooler). $\G_\a$ is an adimensional geometrical factor accounting for the present cylindrical setup: $\G_\a\simeq \a-1$ and thus $E_z^{max}$ diverges for $\a$ reaching unity (as it happens in the rectangular cooler for vanishing thickness b). It is to be noted that this value is independent of $q_{IN}$.

The maximum temperature difference is instead reduced by any finite $q_{IN}$ (see supplementary material) but in the case of a single-staged cooler of a superconducting cable where $q_{IN}=0$ it reaches its maximal value and reads:
\be\label{eq:DTmax}
\D T_{MAX}=\frac{1}{2}Z_a^0T_l^2\frac{\log{\a}}{\G_\a}
\ee
where $Z^0_a$ is the adiabatic figure of merit corresponding to $Z^0\equiv N^2 B_0^2/K\rho$, the isothermal figure of merit at $B_0$, the magnetic field in $r_0$ ($Z^0$ is independent of temperature in the chosen model).

\begin{figure}[h]
\begin{center}
\includegraphics[width=9cm]{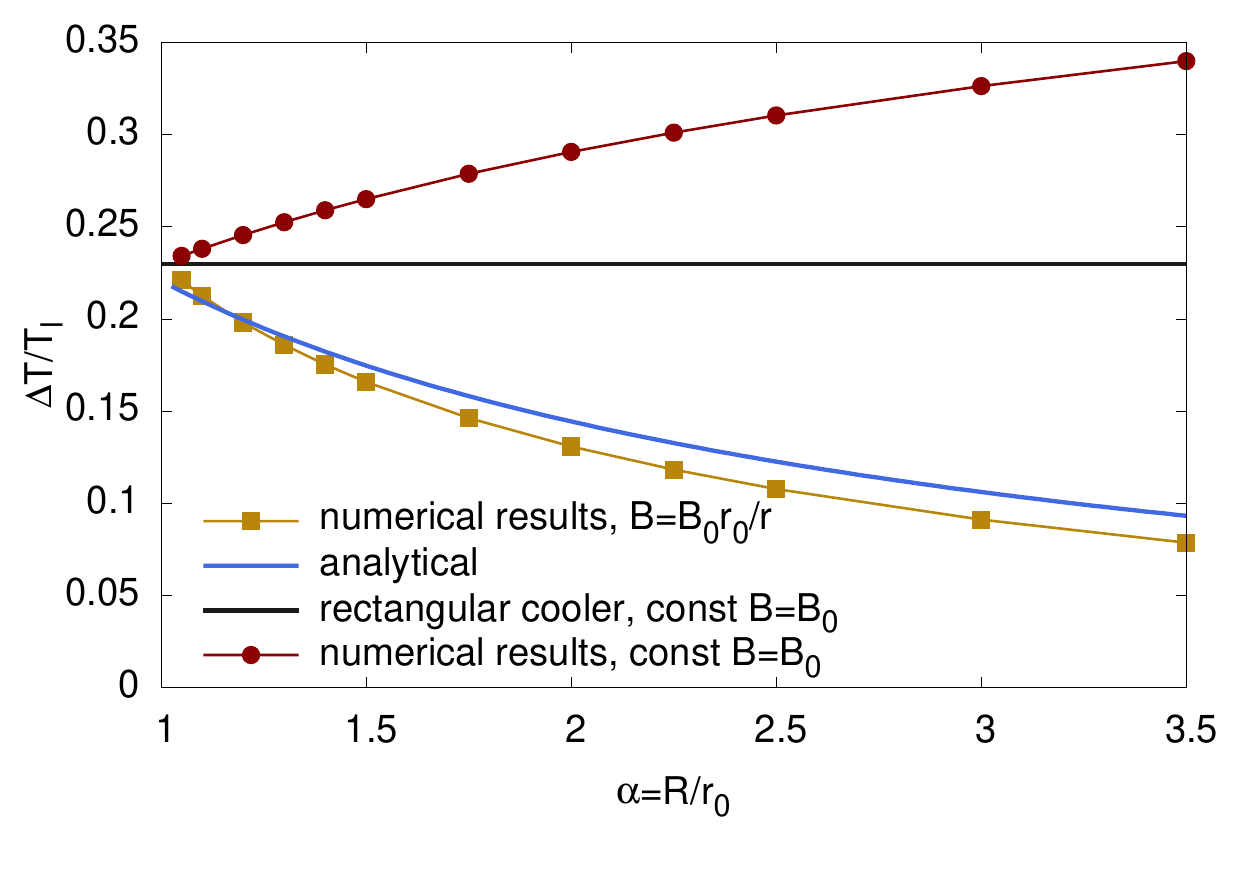}
\caption{Maximum relative temperature drop (i.e. for $E_z=E^{MAX}_z(\a)$) for a model of circular Ettingshausen cooler with constant coefficients ($\rho$=2.5m$\Omega$cm, $K$=0.062W/Kcm, $N$=35$\mu$V/KT, yielding $Z\equiv$0.00125K$^{-1}$ roughly representative of Bismuth at ambient temperature and B$\simeq$12T) for a magnetic field at the interface between the cable core and the thermomagnetic coating of B$_0$=12.6T (e.g. generated by a current of density 10$^5$ A/cm$^2$ running in a core of radius r$_0$=2~cm), plotted as a function of the thickness $\a$. The results  of the numerical solution (squares) and the analytical result eq.(\ref{eq:DTmax}) are plotted, validating the minor approximation used in the analytical treatment (see supplementary material). The red full dots represent the numerical results for the same parameters but for a magnetic field artificially kept constant (B=B$_0$) at all distances. The black line is the result for a rectangular cooler eq.(\ref{eq:slab_DTmax}) at the same B=B$_0$. These results show that the order of magnitude of the temperature drop (for zero heat load $\q_{IN}=0$) is correctly described by formula eq.(\ref{eq:DTmax}) (being $\sim 50K$ for the present parameters), and that, for constant coefficients, in the cylindrical geometry a slow reduction with the material thickness comes from the decay of the magnetic field, that supersedes the advantages of the cylindrical shape over the rectangular one in terms of cooling.}
\label{fig:Cable_fixpar}
\end{center}
\end{figure}

We are then in the position to assess the order of magnitude of the temperature drop that can be obtained by the present cooling system. Indeed Fig. \ref{fig:Cable_fixpar} shows that the analytical model captures very well the result of the numerical solution (lines vs squares), and from formula (\ref{eq:DTmax}) one sees that for a circular cooler of vanishing thickness, since $\lim_{\a\rightarrow 1}\frac{\log{\a}}{\G_\a}=1$, the maximum temperature drop is equivalent to that for a rectangular cooler (in a constant magnetic field $B_0$) eq. (\ref{eq:slab_DTmax}). For parameters representative of Bismuth at very high magnetic fields and ambient temperature (yielding $Z^0_aT\sim 0.4$), $\D T_{MAX}$ is of order $\sim50K$.
When the cylindrical cooler has a finite thickness, however, the value of $\D T_{MAX}$ is reduced by the geometrical factor $\log{\a}/\G_\a$, which is a slowly decaying function of $\a$. This reduction shows that the decay of magnetic field with the distance from the core cable prevails on the aforementioned advantages of the cylindrical shape. 
However, it is remarkable that if a constant $B=B_0$ is fictitiously imposed, then the $\D T/T_l$ (calculated numerically) is on the contrary enhanced for thicker circular coolers (Fig. \ref{fig:Cable_fixpar}, dots). This illustrates explicitly the improvement of cooling power over a rectangular shape (for which $\D T/T_l$ is independent of the thickness, eq.(\ref{eq:slab_DTmax})) brought in by the circular geometry.
A constant magnetic field is indeed artificial in this geometry but, as we will see later on, this advantage can still be exploited in a real cooler whenever the figure of merit of the thermomagnetic material becomes insensitive to the decay of the magnetic field. There, the circular geometry yields enhanced performances.

We have not considered thus far the overall efficiency of the proposed cooler. 
Indeed energy will be dissipated to maintain the temperature gradient and heat will be expelled (even when $q_{IN}=0$), in accord with the second principle of thermodynamics.
In the case in which $q_{IN}=0$ it is useless to consider the customary coefficient of performance or COP$=q_{IN}/W$, where W is the expenditure of electric power (per unit cable length) for the cooling $W=2\pi\int^R_{r_0} rdr E_z j_z(r)$, since the COP vanishes. We will just consider $W=Q_{OUT}=2\pi R q_r(R)$ (where we used eq. \ref{eq:continuity}), i.e. the total expelled heat for unit cable length as the measure of the convenience of the cooling process, for a given temperature drop. 
In the considered model (see the formula in the supplementary material) for vanishing thickness $Q_{OUT}$ recovers exactly the result for rectangular coolers and diverges in a way inversely proportional to the thickness, in accord with the divergence of $E^{MAX}_z$.
For finite thickness $Q_{OUT}$ decays (like $\log^2{\a}/\a^2$ at large $\a$). It should also be noted that $Q_{OUT}$ depends only on the ratio $\a=R/r_0$, but for a larger device (i.e. larger $r_0$ and R, at constant $\a$) the same heat is distributed on a larger surface (i.e. $q_r(R)$ is smaller) so it will be more convenient for the thermalization to the ambient temperature.

Thus in the present model with constant coefficients a compromise in choosing the coating thickness, between the decay of $\D T$ and that of $Q_{OUT}$, has to be found, i.e. a regime in which the amount of power needed to reach the desired temperature drop doesn't make the whole process largely disadvantageous compared to other cooling systems. In this view, as shown for Peltier cooling in Ref. \onlinecite{Shilliday_circ_peltier}, the main practical advantage of a circular over a rectangular setup is the automatic thermal insulation ensured by the circular geometry, and absence of border effects far from the cable ends.

However turning to more realistic calculations we will see that whenever the material figure of merit becomes insensitive to the magnetic field the circular geometry has a neat advantage in terms of higher temperature drops.

\medskip

{\bf \footnotesize \noindent NUMERICAL RESULTS FOR A REALISTIC SETUP REALIZED IN BISMUTH}

\noindent Indeed a simulation for a cable cooler using the realistic properties of metallic Bismuth is possible. Using available data in the literature, a map of $\rho(T,B)$\cite{Kapitza_Res_Bi},$K(T,B)$\cite{kaye_ThCond_Bi,Uher_Goldsmid_Comp_thermomag_mat}\footnote{For thermal conductivity, data at large B and at low T have been extrapolated from the available literature and in comparison with some unpublished data from Collaudin et al. The very roughly extrapolated data used here remains however on the conservative side for thermomagnetic performances, as K never gets below 0.05 W/Kcm, which is likely to be an overestimate of the thermal conductivity at high fields at room temperature\cite{goldsmid_book2009}.} 
and $N(T,B)$
\footnote{In the absence of as extended (in both B and T) data as needed for the Nernst coefficient, data for the isothermal figure of merit from Ref. \onlinecite{harman-ZNE_Bismuth} were used along with the data for $\rho$ and K to deduce N. Since K is likely to be somewhat overestimated at large fields, the calculated N could be slightly overestimated too, not affecting the overall performance however, since ZT is kept fixed.} 
along the crystal axes of higher performance\footnote{The crystal directions for which the Ettingshausen effect is maximized in Bi and its Sb alloys are: current along the trigonal symmetry axis, temperature gradient along the binary axis and magnetic field along the bisectrix axis\cite{harman-shaping101K}.} can be traced in the ranges B=0$\div$13T and T=70$\div$300K, suitably interpolating to obtain the values at any B,T, which is needed in order to numerically solve the system of equations (\ref{eq:Jz}),(\ref{eq:qr}) and (\ref{eq:continuity}). 

The maximum temperature drop that can be reached within the realistic simulations as a function of the thickness is shown in Fig. \ref{fig:Cable_Bi}, along with the amount of expelled heat. 
\begin{figure}[h]
\begin{center}
\includegraphics[width=9cm]{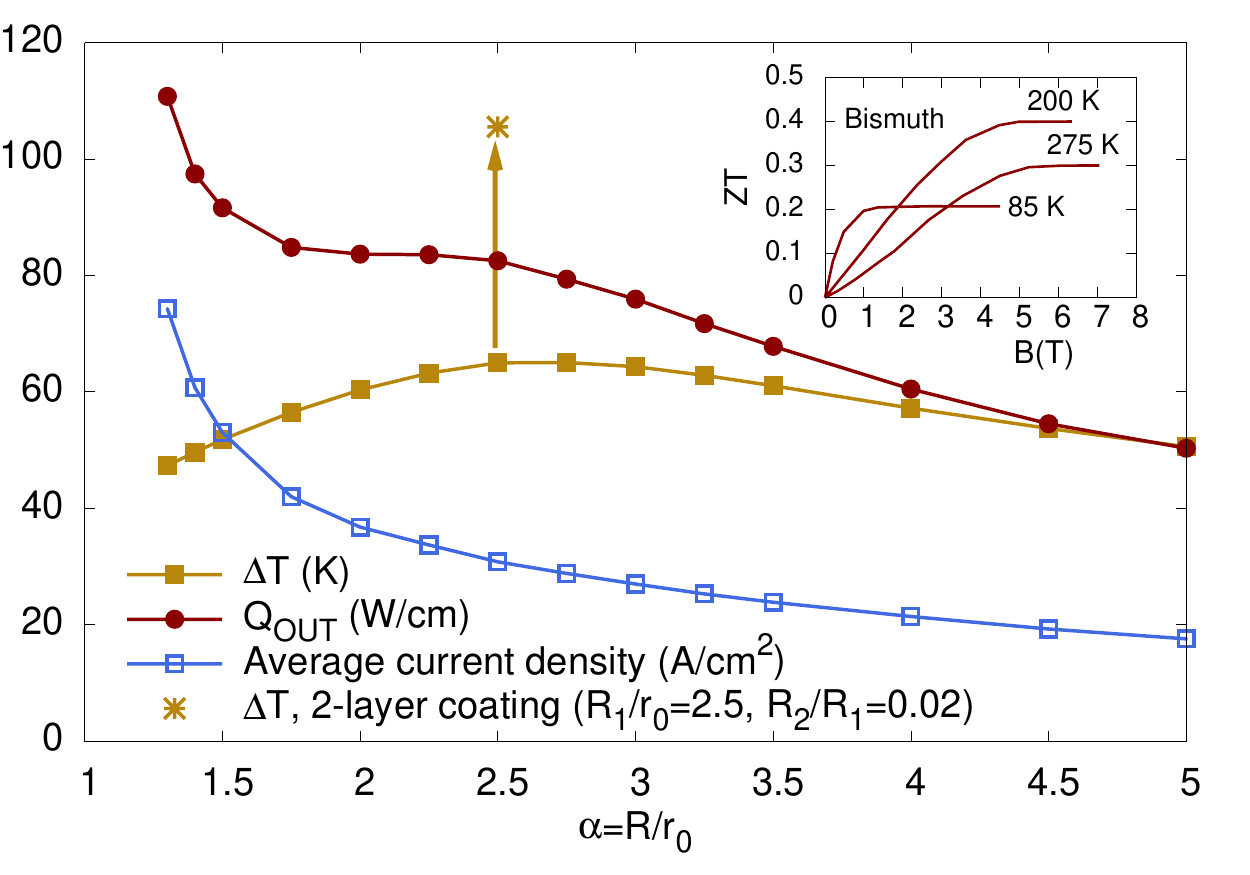}
\caption{Numerical simulations at maximum cooling (i.e. for $E_z=E^{MAX}_z(\a)$) for a circular Ettingshausen cooler using realistic material properties for single-crystal Bismuth along the crystal axes of maximum thermomagnetic performance, and a magnetic field equal to B$_0$=12.6T at the interface between the cable core and the thermomagnetic coating (generated by a current of density 10$^5$ A/cm$^2$ running in a core of radius r$_0$=2~cm). Data in the main plot show (in units as indicated in the legend) the temperature drop $\D T$, the total expelled heat for unit cable length and the average current density running in the thermomagnetic coating, as a function of the thickness of the cooling layer $\a=R/r_0$. The improvement for one realization of a double-layer coating (see text) to $\D T$=105.5K is indicated by the arrow. {\bf Inset}: measured figure of merit for Bi (reproduced from \onlinecite{harman-ZNE_Bismuth}), showing the saturation field and saturation value for ZT at various temperatures. As long as the cooler is immersed in high enough magnetic field so that the thermomagnetic figure of merit assumes its saturation value, $\D T$ raises with the cooler thickness.}
\label{fig:Cable_Bi}
\end{center}
\end{figure}

The electric field $E^{MAX}_z$ giving the maximum temperature drop for each $\a$ is very well described by the analytical expression for constant material parameters eq. (\ref{eq:Emax}), while the $\D T$ is seen to raise with $\a$ (in contrast with the constant coefficient model considered above) below $\a\simeq 2.75$, where it starts decreasing. This behavior is easily interpreted by looking at the measured figure of merit for this material\cite{harman-ZNE_Bismuth} (inset of Fig. \ref{fig:Cable_Bi}). Indeed it is known\cite{goldsmid_book2009} that when the condition $\mu_eB>>1$ is met, where $\mu_e$ is the electron mobility, the figure of merit saturates. This happens obviously at higher fields, for higher temperatures, at which the electron mobility is lower. It is seen that $B\sim 5T$ is the saturation field at room temperature (implying also that at all lower temperatures ZT will have already reached its saturated value). Indeed in our setup, if the whole thermomagnetic layer is immersed in a field above the saturation value at room temperature, then its properties will be insensitive to changes in B, and $\D T$ is expected to grow, as shown in Fig. \ref{fig:Cable_fixpar}. The magnetic field is the lowest at the external border (i.e. $r=R$) and it diminishes with growing $\a$, so as long as $\a$ is such that $B(r=R)$ is above 5T we expect $\D T$ to grow, and only for larger $\a$'s to start the logarithmic decrease expected in the analytical model. It is easily checked that $B(R)=B_0/\a\simeq 5T$ when $\a\simeq 2.5$, for the chosen $B_0=12.6T$, so the rationale matches the simulations.  

As a partial conclusion then, numerical simulations for realistic parameters show that the proposed Ettingshausen cable cooler realized in pure Bismuth may in principle reach more than 60K temperature drop from the ambient temperature if the current running through the core is strong enough to generate a field at the interface of $\sim12T$, which is a realistic value for a superconducting cable (and $\D T$ would grow even more for higher magnetic fields and higher $\a$'s). This would be done with moderate energy cost, as the expelled heat would be of $\sim$50-80 W/cm of cable length, which is a very encouraging result.

Besides the evaluation of the heat loss however, one realizes that higher temperature drops are needed in the hope of using this cooling system down to high-Tc superconductor critical temperatures. The usual way to go to do this is using multiple cooling stages\cite{kooi_staging}. 

\medskip

{\bf \footnotesize \noindent STAGING}

\noindent Indeed in this context multiple layers of (possibly different) thermomagnetic materials, each with a different electric field (and thus a different current running through them) can further improve the performance.
Adding a second stage yields a temperature drop $\D T={\D T}_1+{\D T}_2$ which is a complicated function of the two electric fields and the two radiuses $E_{z1},E_{z2},R_1,R_2$. In particular ${\D T}_2$ will be reduced due to the heat flux $q_{IN2}=q_{OUT1}$ expelled from the inner stage. An exact theory of staging optimization is beyond the scope of this article, however the generalization of the formula for $\D T$ eq. (\ref{eq:DTmax}) to finite $q_{IN}$ (see supplementary material) suggests a simple way to nearly double the $\D T$. Indeed ${\D T}_2$ is reduced by a term $\propto q_{IN2}R_1\log{\a_2}$, that can thus be minimized by minimizing the thickness of the layer $\a_2=R_2/R_1$, which also maximizes the term proportional to $Z^0_a$ since the geometrical factor in eq. (\ref{eq:DTmax}) tends to 1. Thus in principle a double coated cable with the external cooler layer much thinner than the internal one, in which both electric fields would be tuned to the respective $E^{MAX}_z$, would nearly double the $\D T_{MAX}$.

This can be successfully simulated and as an example the configuration (keeping the realistic material properties for Bi and the same core current and radius $r_0=2cm$ as before) with R$_1$=5cm, R$_2$=5.1cm, E$_{z1}$=0.065V/cm and E$_{z2}$=0.48V/cm yields $\D T$=105.5K (see Fig.\ref{fig:Cable_Bi}). This excellent result is however reached at the expenses of a very large energy consumption ($Q_{OUT}$=1072W/cm!), since E$_{z2}$ is very large to match the value for maximum cooling in the very thin second layer. Indeed a compromise between temperature drop and heat losses is part of the optimization process and is beyond the scope of this article. Nevertheless we judge the present results of $\D T> 100K$ very encouraging for future applications.

\medskip

{\bf \footnotesize \noindent DISCUSSION: MAIN OBSTACLES AND IMPROVEMENT OF MATERIALS}

\noindent At this point of the discussion, a few obstacles need to be pointed out. One obvious problem in the case of the cooling of a superconducting cable is the starting procedure. Indeed in the steady state the presented setup runs in a "symbiotic" configuration: the superconducting current provides the high magnetic field needed to operate the cooler, which in turn provides the low temperature needed for the superconductor to work. This mode of operation seems reachable only if another, more conventional cooling system is used for initiating the steady state\footnote{Less likely, one could use the normal metallic part of the wire in the core usually intertwined with the superconductor to briefly provide the strong magnetic field. But it seems difficult that this can be realized long enough for the cooling to operate until the realization of the superconducting state, without instead just overheating.}.

Another point to be considered is that for the device to actually work with a superconductor the temperature to be reached is well below $T_c$. 
Indeed T$_c$ is lowered by a finite magnetic field and current density. These three parameters are interdependent have to be inside a zone in the parameter space (see e.g. \cite{Chaddah_Ic_supercond} for Nb$_3$Sn) such that all are far from their highest values.  
Thus the needed temperature drops are even larger than just reaching T$_c$(B=0, i=0) and have to be obtained with magnetic fields well below the critical values, etc.

A third point is that Bismuth, today's champion thermomagnetic material at ambient temperature, has very anisotropic properties. Since it is probably unpractical to coat a cable in a layer of oriented single-crystals of Bismuth such that $ZT$ is maximized at all angles, the actual performance is likely to be inferior to what the simulations presented in this article show.

Lastly, for applications at temperatures below 100K (e.g. in a possible setup in which a low-T$_c$ superconductor is cooled with the thermomagnetic coating in order to operate in an external bath of liquid Nitrogen, much less expensive than the normally used liquid Helium) unfortunately Bismuth is not useful. Indeed its thermomagnetic properties degrade quickly below  $\sim$100K. \cite{Behnia-Bismuth}

All the listed hindering point could be overcome however if better thermomagnetic performance than in Bismuth and its alloys\cite{yim1972bi,jandl1994thermogalvanomagnetic,lenoir1996transport} could be obtained. 
Indeed thermoelectric materials have witnessed enormous developments in the last decades, while thermomagnetic materials have received much less attention. Few exceptions are noteworthy, and encouraging. For instance lately the large value of $Z_aT$=0.5 in the range 35-50K has been reported for Li$_{0.9}$Mo$_6$O$_{17}$\cite{cohn_Li0.9Mo6O17}.

The guidelines for the search of efficient thermomagnetic materials have been laid down\cite{delves1962prospects,horst1963thermomagnetic,Behnia_GiantNernstMaterials,Behnia-Nernst_FermiLiquid}: ambipolarity (i.e. the presence of both electron and hole carriers), high-mobilities, low Fermi energies (i.e. shallow bands).
These conditions are difficult to be optimized altogether, however heavy-fermionic materials (in which the electronic masses enhanced by the strong electronic correlations favor low Fermi energies) are being explored\cite{Behnia_GiantNernstMaterials} and some of them like URu$_2$Si$_2$ and PrFe$_4$P$_{12}$ have shown a large Nernst effect at temperatures of a few Kelvins, holding promises for Ettingshausen cooling in that range of temperatures. Other materials show large Nernst signals like correlated semiconductor FeSb2\cite{Sun-FeSb2_Nernst}, Iron pnictides (see e.g. C. Hess in \cite{Zlatic_Hewson-ThermoelectricMat_Book}) and others, but the optimization of thermomagnetic materials remains essentially uncharted territory as of today. Very rough estimates let however imagine that nothing prevents reaching ZT as high as 0.8,\cite{harman-InfiniteStaged_theory} in principle, which would yield exceptional performances.

\medskip

{\bf \footnotesize \noindent CONCLUSIONS}

\noindent In conclusion, a solid-state mechanism for building self-cooling high-power cables was proposed in this article. An analytical treatment of the differential equations for the temperature and heat current distributions was given for the case of constant material parameters. The estimate for the maximal temperature drop that can be obtained in this setup is similar - actually slightly inferior for finite thickness - to that of a standard rectangular cooler (in a constant magnetic field equal to the maximum field in the circular cooler B$_0=B(r_0)$), because the advantage in heat pumping brought in by the circular shape is overrun by the decay of the magnetic field with distance. However remarkably numerical simulations with realistic parameters for Bismuth coatings show that for a coating layer up to 2.5 times wider than the core the temperature drop from ambient temperature increases until $\simeq$60K. This is due to the saturation of the thermomagnetic properties in high magnetic fields, so that as long as the entire coating is immersed in a field larger than 5T for Bismuth the performance is insensitive to the decay of B(r) and only the advantage due to the geometrical properties of the circular setup remain.
Finally results from simulations for a 2-layer staged cooler are given, showing temperature drops larger than 100K (even if with a much larger electrical expenditure) from the ambient, as indeed predicted by the analytical model. 

These results are encouraging for applications such as high-power cables for long-distance transport of electric power, but also with some adjustments to the cooling of magnet coils, among others. An improvement on the present-day performance of the known thermomagnetic materials is probably needed for any of these applications to become technologically convenient, and hopefully this article will stimulate further research in this subfield, which has received much less attention than its thermoelectric counterpart, thus far.

\acknowledgements

The author is indebted to P. Bruno, V. Corato, L. De Leo, G. Lang, J. Lesueur, M. Schir\`o and P. Stoliar for precious discussions, as well as to K. Balaa, A. Hassine and J. Lewiner, also for their concrete support to this project, to A. Collaudin for sharing some of her unpublished data and to A. Georges for driving the author's interest to the field with Ref. \onlinecite{goldsmid_book2009}. 

A French patent request is pending under N. 15 51663 for the setup and mechanism treated in this article.

\clearpage


\begin{widetext}
\begin{center}
{\bf\large A thermomagnetic setup for self-cooling cables: Supplementary information}
\end{center}
\vspace{0.5cm}
\end{widetext}

\renewcommand{\thepage}{S\arabic{page}}  
\renewcommand{\thesection}{S\arabic{section}}   
\renewcommand{\thetable}{S\arabic{table}}   
\renewcommand{\thefigure}{S\arabic{figure}}

\setcounter{page}{1}
\setcounter{figure}{0}
\setcounter{table}{0}

\section{Isothermal and adiabatic figures of merit}

All the physical properties of a thermomagnetic material used in the main text (the resistivity $\rho$, the Nernst coefficient $N$, the thermal conductivity $K$) are intended defined in a transversely isothermal setup (e.g. the isothermal resistivity is measured keeping the transverse temperature gradient to zero). The figure of merit $Z=\frac{(NB)^2}{\rho K}$
is thus called the \emph{isothermal} figure of merit.
It is also useful to define the \emph{adiabatic} figure of merit $Z_a=\frac{(NB)^2}{\rho_a K}$, which is an analogous quantity in which appears the adiabatic resistivity. The latter is measured keeping the transverse heat flow to zero and reads\cite{callen2006thermodynamics,HarmanHonig-I} $\rho_a=\rho(1-ZT)$. 

The two figures of merit are related by $Z_a=\frac{Z}{1-ZT}$, so that $Z_aT$ is always larger than ZT and their range of physical values is different in that $0<ZT<1$ while $0<Z_aT<\infty$.\cite{kooi-Ettingshausen_cooler}

Measured ZT for Bi and BiSb alloys reaches, in high magnetic fields, values between 0.2 and 0.4 in the temperature range 100-300K, as shown in the inset of Fig. 3 in the main text\cite{harman-ZNE_Bismuth}. Correspondingly $Z_aT$ assumes values between 0.25 and 0.66.

The adiabatic figure of merit $Z_a$ plays in thermomagnetic applications the role that the isothermal thermoelectric figure of merit $Z_{SP}=\frac{S^2}{\rho K}$ (where S is the Seebeck coefficient) plays in thermoelectric (Seebeck-Peltier) applications\cite{kooi-Ettingshausen_cooler, goldsmid_book2009}.

\section{Analytical treatment of the circular Ettingshausen cooler with constant coefficients}

The fundamental equations for the heat flow and temperature distribution for a circular Ettingshausen cooler, as we have seen in the main text are:
\bea
\label{eq:Jz} j_z&=\#&\frac{E_z}{\rho}+\frac{NB}{\rho}\frac{dT}{dr},\\
\label{eq:qr} q_r&=\#&\frac{NBT}{\rho}E_z+K(ZT-1)\frac{dT}{dr},
\eea
where $j_z$ and ${E_z}$ are the electric current density and the electric field respectively (${\mathbf E}=-{\mathbf \nabla} \mu$, where $\mu$ is the electrochemical potential), along the direction of the cable , whereas $q_r$ is the heat current density in the radial direction and $T(r)$ is the absolute temperature. $N$, $\rho$ and $K$, and thus $Z$ are in general functions of temperature and magnetic field, and indeed material-dependent.

All quantities are constants in $z$ and $\theta$ because of the cylindrical symmetry but are indeed functions of $r$. However the continuity of the elettrochemical potential and the symmetry-imposed constant value of $E_r$ as a function of z implies that $E_z$ is a constant in r, because\cite{kooi-Ettingshausen_cooler} 
$${\partial E_z}/{\partial r}=-\frac{\partial^2 \mu}{\partial r \partial z}={\partial E_r}/{\partial z}=0.$$ 
This implies in particular, through eq.~(\ref{eq:Jz}), that the  current density $j_z$ likely varies with r.

$B(r)= B_0r_0/r$ is the intensity of the magnetic field (oriented along $\hat \theta$), which we parametrize by its value at the interface between the conducting core and the thermomagnetic coating 
$
B(r=r_0)=B_0=\frac{\mu_0}{2\pi}\frac{I}{r_0}=\frac{\mu_0}{2}r_0i,
$  
where $\mu_0$ is the vacuum permittivity and $I$ and $i$ are the total current and current density in the cable core, respectively
(in principle one should also consider the magnetic permittivity of the thermomagnetic material $\mu_r$ in the calculation of B. However even in strong diamagnetic materials (such as Bi) $\mu_r$ is in practice 1).
The current density i is that of the main current transported by the cable, running in the cable core and generating the magnetic field, and should not be confused with the density of auxiliary current $j_z$, which runs in the thermomagnetic coating and is responsible for the Ettingshausen cooling effect. Typically $i$ will be several orders of magnitude larger than $j_z$. 

Equations (\ref{eq:Jz}) and (\ref{eq:qr}) only express the heat and charge current response of a material submitted to a temperature and electrochemical potential gradient at a given point, but in order to find the temperature and current distribution energy conservation has to be enforced at each point by a continuity equation\cite{kooi-Ettingshausen_cooler} $\nabla \cdot ({\mathbf q} + \mu {\mathbf j})=0$, that here specializes to:
\be\label{eq:continuity}
\frac{1}{r}\frac{d\;\: rq_r}{dr}=E_zj_z
\ee

We will now solve the case with constant $\rho$, $N$ and $K$ (and thus $Z$).
Indeed equations (\ref{eq:Jz}) and (\ref{eq:continuity}) can be combined and, in the case of constant $\rho$ and $N$, solved for $q_r(r)$ giving:
\be\label{eq:sol_qr}
q_r(r)= \frac{NB_0r_0}{\rho}E_z\frac{T(r)}{r}+ \frac{E_z^2}{2\rho}r+\frac{c_0}{r}
\ee
with $c_0$ a constant in r. 
Also, from (\ref{eq:qr}) and (\ref{eq:sol_qr}) one obtains the equation for the temperature distribution $T(r)$:
\be\label{eq:T_r}
K(ZT-1)\frac{dT}{dr}=\frac{E_z^2}{2\rho}r+\frac{c_0}{r}.
\ee

A difficulty here is that Z depends on r through the magnetic field $B(r)\sim 1/r$. In order to proceed analytically we make, in the left hand side of eq. (\ref{eq:T_r}), the approximation $Z(r)T(r)=\overline{ZT}$, where $\overline{ZT}$ is a constant in r. This may be not too bad of an approximation, since in our setup Z(r) diminishes with r and T(r) grows. Moreover in realistic situations - and this even prescinds from specific geometrical considerations -, at large values of the magnetic field ZT is known to saturate, becoming constant in B, while an optimal choice of the thermomagnetic material in the range of temperatures of interest implies that ZT should be near to its maximum and thus likely quite constant for a range of temperatures. As a practical example - as shown in the main article (inset of Fig. 3) - for fields above 5T and temperatures between 100K and 300K Bismuth has a rather constant value of ZT between 0.2 and 0.4\cite{harman-ZNE_Bismuth}. However it does so thanks to a compensation of the dependence of $\rho(T,B)$, $N(T,B)$ and $K(T,B)$, which is not taken into account here since we have assumed these quantities to be constants, thus our treatment remains approximate in any case.

Solving eq. (\ref{eq:T_r}) with this approximation (and also considering a constant K) leads to the expression for the temperature dependence:
\be\label{eq:T_r_sol}
T(r)=\frac{\frac{E_z^2}{4\rho}r^2+c_0\log{r}+c_1}{K(\overline{ZT}-1)},
\ee
with $c_1$ another constant in r. 

Two kind of boundary conditions can be used to parametrize the solutions through $c_1$ and $c_0$. Indeed the external (heat sink) temperature is fixed for a thermostated device, i.e. $T(R)=T_h$, and either the internal temperature $T(r_0)=T_l$ or the heat flow at the source $q_r(r_0)=q_{IN}$ can be fixed (in each case the other quantity becoming a function of the fixed one). It is the latter that is useful in the context of this article. When fixing the boundary conditions so that $T(R)=T_h$ and $q_r(r_0)=q_{IN}$ one obtains the two conditions
\bea \label{eq:boundary_q_in}
c_0&\!=&\!q_{IN}r_0-\frac{E_z^2}{2\rho}r_0^2-E_z\frac{NB_0T_l}{\rho}r_0\\ \nonumber
c_1&\!=&\!K(\overline{ZT}-1)T_h-\frac{E_z^2}{4\rho}R^2-c_0\log{R}
\eea
(in these implicit expressions $T_l=T_l(T_h, q_{IN})$ through eq. (\ref{eq:T_r_sol}) and (\ref{eq:boundary_q_in})).

The alternative parametrization $T(R)=T_h$ and $T(r_0)=T_l$ (not used in this article) yields $c_0=[(T_h-T_l)K (\overline{ZT}-1)-\frac{E_z^2}{4\rho}(R^2-r_0^2)]/\log{(R/r_0)}$ and $2c_1=(T_h+T_l)K(\overline{ZT}-1)-\frac{E_z^2}{4\rho}(R^2+r_0^2)-c_0\log{Rr_0}$.

Indeed for a cooling device coating a superconducting core (the latter not producing any heat in the conduction) $q_{IN}=0$ in the steady state, whereas for the device coating a resistive core $q_{IN}$ is imposed and finite. Also for all stages beyond the first in a multi-stage setup $q_{IN}$ is finite.

We are interested here in knowing what is the maximum temperature drop that one can obtain for a given $q_{IN}$. The electric field that minimizes $T_l$ at fixed $q_{IN}$ - which is given by the same condition that maximizes the cooling power $q_r(r_0)$ for any fixed $T_l$ - is:
\be\label{eq:Emax_supp}
E_z^{max}=\frac{NB_0T_l}{r_0\G_\a},
\ee
where $\G_\a=\frac{\a^2-1}{2\log{\a}}-1$, and $\a=R/r_0$ measures the thickness of the thermomagnetic layer ($\a>1$).
This formula is very similar to the case of a rectangular cooler\cite{kooi-Ettingshausen_cooler} (where $E_z^{max}={NBT_l}/{b}$, and $b$ is the thickness of the cooler). $\G_\a$ is a geometrical factor accounting for the present cylindrical setup: $\G_\a\simeq \a-1$ and thus $E_z^{max}$ diverges for $\a$ reaching unity (as it happens in the rectangular cooler for vanishing thickness b).
The expression eq.(\ref{eq:Emax_supp}) is exact in our model, i.e. it does not depend on the approximation $ZT=\overline{ZT}$, because the terms involving $\overline{ZT}$ disappear when performing the derivative in $E_z$ (even without doing the aforementioned approximation this holds true because the derivative of the terms in question vanish at the extremal point).

The maximum temperature difference then reads:
\be\label{eq:DTmax_qIN}
\D T_{MAX}=\frac{1}{2}\frac{Z^0}{1-\overline{ZT}}T_l^2\frac{\log{\a}}{\G_\a}-q_{IN}\frac{r_0\log{\a}}{K(1-\overline{ZT})}
\ee
where $Z^0\equiv N^2 B_0^2/K\rho$ is the isothermal figure of merit corresponding to $B_0$, the magnetic field in $r_0$ (and is independent of temperature in the chosen model). 

In the case of a single-staged cooler of a superconducting cable $q_{IN}=0$, and taking into account that $\frac{Z^0}{1-\overline{ZT}}\simeq Z^0_a$ (see below),
the adiabatic figure of merit in $r_0$, the formula eq. (\ref{eq:DTmax_qIN}) then reduces to the result:
\be\label{eq:DTmax_supp} 
\D T_{MAX}=\frac{1}{2}Z_a^0T_l^2\frac{\log{\a}}{\G_\a}
\ee

In this compact formula, useful for direct comparison with the analogous one for rectangular coolers eq. (1) of the main article, we have however used here $Z_a^0$ in a quite sloppy way. Indeed even in this simplified model, rigorously the adiabatic figure of merit reads $Z_a^0(T)=Z^0/(1-Z^0T)$ and thus it is a function of temperature. Then, depending on the precise choice of $\overline{ZT}$ in our approximated treatment, $\frac{Z^0}{1-\overline{ZT}}$ can coincide with $Z^0_a(\tilde T)$ (if one takes $\overline{ZT}=Z^0\tilde T$, with $\tilde T$ any reasonable temperature, e.g. $T_l$) or can assume other values.

\begin{figure}[h]
\begin{center}
\includegraphics[width=9cm]{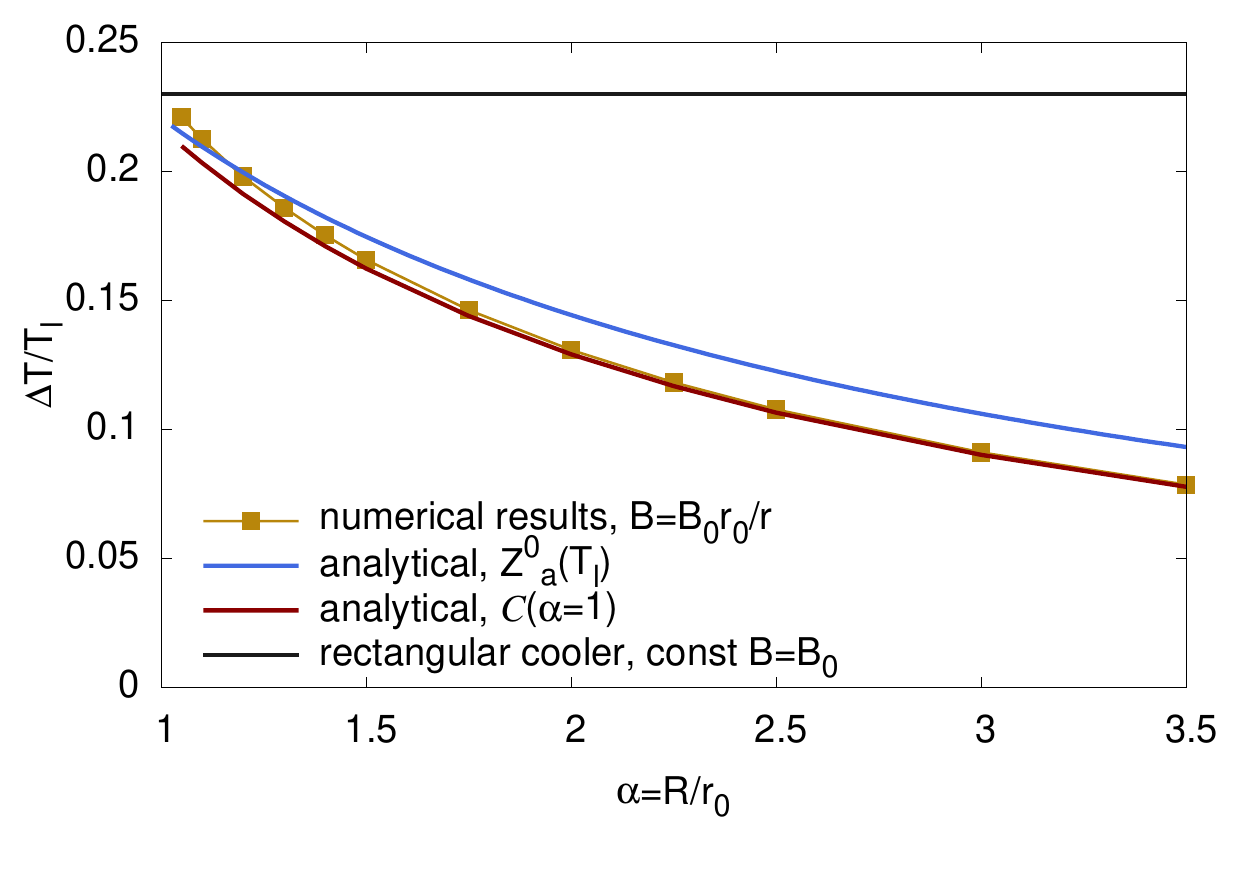}
\caption{Maximum relative temperature drop (i.e. for $E_z=E^{MAX}_z(\a)$) for a model of circular Ettingshausen cooler with constant coefficients (same parameters than in Fig. 2 of the main article) plotted as a function of the thickness $\a$. Besides the results of the numerical solution (squares), the analytical result eq. (\ref{eq:DTmax_supp}) is plotted, for two typical choices of $\overline{ZT}$, as explained in the text. These results show that the order of magnitude of the temperature drop (for zero heat load $\q_{IN}=0$) is correctly described by formula (eq. (\ref{eq:DTmax_supp})), and illustrate the reduction with material thickness in the cylindrical geometry compared to the rectangular one at fixed $B=B_0$, due to the decay of the magnetic field, that supersedes the advantages of the cylindrical shape, in this model with fixed coefficients.}
\label{fig:Cable_fixpar_supp}
\end{center}
\end{figure}

Indeed the precise choice of $\overline{ZT}$ for any given $\a$ in the analytical model is somewhat arbitrary in the present treatment, however most reasonable choices yield an analytic $T_l$ fitting the numerical data (at least for the physical parameters chosen here) within a few \% over a range $\a$=1$\div$3.5. This is shown by the comparison of two such choices in Fig. \ref{fig:Cable_fixpar_supp} (solid lines), and their nice agreement with the exact numerical solution of the model (dots).
This justifies the use of a "generic" $Z_a^0$ that we have made in the main article, however one should keep in mind this point. 

Examples of possible choices are the natural mean-field-like choice $\overline{ZT}= < Z(r)T(r) > \simeq < Z(r) >< T(r)>$ and the choice $\overline{ZT}=Z^0T_l$, leading to $\frac{Z^0}{1-\overline{ZT}}=Z^0_a(T_l)$. In the latter case, solving explicitly for $T_l$ gives:
\be
T_l=\frac{1+Z_0T_h-\sqrt{(1-Z_0T_h)^2+4Z_0T_h\frac{\log{\a}}{2\G_\a}}}{2Z_0(1-\frac{\log{\a}}{2\G_\a})}
\ee
The blue solid line (titled $Z^0_a(T_l)$) plotted in Fig.\ref{fig:Cable_fixpar_supp} is $\D T/T_l$ calculated using this expression and shows a good agreement with the numerical solution.

The choice $\overline{ZT}=Z^0\bar T$ (where $\bar T = (T_h+T_l)/2$ is the middle temperature in the cooler) is instead further off at large $\a$ but reproduces exactly the $\a=1$ limit, conciding with the result for a rectangular cooler (eq.(1) of the main article, black solid horizontal line in the Figure). Much more precise results reproducing the dependence of $\D T_{MAX}$ on $\a$ (particularly for large $\a$) are obtained setting (in eq. (\ref{eq:DTmax_qIN}) for $q_{IN}=0$) the fraction $\frac{Z^0T_l}{1-\overline{ZT}}= \mathcal{C}=\frac{Z^0T_l(\a=1)}{1-Z^0T_l(\a=1)},$ i.e. to a constant in $\a$ in which the value of $T_l$ is replaced with that obtained for vanishing thickness. This gives $T_l=\frac{T_h}{1+\frac{\mathcal{C}}{2}\frac{\log{\a}}{\G_\a}}$, which is used for calculating the red solid line (titled $\mathcal{C}(\a=1)$) plotted in Fig.\ref{fig:Cable_fixpar_supp}.

All in all, as said all the mentioned choices yield very similar values of $\D T_{MAX}$  from the viewpoint of a semi-quantitative discussion, thus in the main article we have used use the simple expression $Z^0_a\simeq\frac{Z^0}{1-\overline{ZT}}$, 

When assessing the order of magnitude of the present cooling system, Fig. \ref{fig:Cable_fixpar_supp} shows that the analytical model captures very well the result of numerical simulations (lines vs squares), and from formula (\ref{eq:DTmax_supp}) one sees that for a circular cooler of vanishing thickness, since $\lim_{\a\rightarrow 1}\frac{\log{\a}}{\G_\a}=1$, the maximum temperature drop is equivalent to that for a rectangular cooler (in a constant magnetic field $B_0$) eq. (1) of the main article. For parameters representative of Bismuth at very high magnetic fields and ambient temperature (yielding $Z^0_aT\sim 0.4$), $\D T_{MAX}$ is of order $\sim50K$.

When the cylindrical cooler has a finite thickness, however, the value of $\D T_{MAX}$ is reduced by the geometrical factor $\log{\a}/\G_\a$, which is a slowly decaying function of $\a$. This reduction shows that the decay of magnetic field with the distance from the core cable prevails on the advantages of the cylindrical shape (as illustrated numerically in the main article, Fig.2, where it is shown that for a constant magnetic field $B=B_0$, $\D T_{MAX}$ grows with $\a$). 

In the main article, for the case in which $q_{IN}=0$,  we have also considered, as a measure of the convenience of the cooling process, the expenditure of electric power (per unit cable length) for the cooling, $W=2\pi\int^R_{r_0} rdr E_z j_z(r)$.  
The continuity equation, eq. (\ref{eq:continuity}) implies $W=Q_{OUT}=2\pi R q_r(R)$ for a given temperature drop. The expelled heat flux (i.e. per unit of the external cable area surface), for $E_z=E^{max}_z$ reads:
\be\label{eq:exp_heat}
q_r(R)=\frac{Z^0T_l}{R\G_\a}K(\frac{T_l}{2}\frac{\a^2-1}{\G_\a}+\D T)
\ee

From this expression one sees that for vanishing thickness - the situation giving the maximum temperature drop - the expelled heat diverges.
Indeed when $R\rightarrow r_0$ i.e. $\a\rightarrow 1$ the term in parenthesis tends to $T_h$ and $\G_\a\sim \a-1$, eq. (\ref{eq:exp_heat}) recovers again exactly the result for rectangular coolers and diverges in a way inversely proportional to the thickness, in accord with the divergence of $E^{MAX}_z$.
For finite thickness $q_r(R)$ decays, and faster than for the rectangular cooler ( where it decays inversely proportional to the thickness) due to the geometrical factor $\a^2/\G_a^2\sim 4\log^2\a/\a^2$ multiplying $1/R$.
However the decay of the expelled heat per unit surface $q_r(R)$ matters only for the removal of heat by the external environment in order to maintain the cable surface temperature $T_h$ (which can also be optimized simply using a larger overall dimension of the setup, thus a larger $r_0$ at constant $B_0$. The actual amount of energy loss (per unit cable length) in the cooling process, i.e. $Q_{OUT}=2\pi R q_r(R)$,  decays in a slower fashion instead (still $\sim \log^2\a/\a^2$ however).
In this view, as shown for Peltier cooling in Ref. \onlinecite{Shilliday_circ_peltier}, the main practical advantage of a circular over a rectangular setup is the automatic thermal insulation ensured by the circular geometry, and absence of border effects far from the cable ends.

\end{document}